\documentclass{ws-ijmpd}
\usepackage{amssymb}
\usepackage{amsmath}
\usepackage{graphicx}
\usepackage{epsfig}
\usepackage{bm}
\begin{document}

\markboth{Sergei M. Kopeikin}
{Comments on the paper by S. Samuel "On the speed of gravity and the Jupiter/Quasar measurement"
}

%
\catchline{}{}{}{}{}
%

\title{Comments on the paper by S. Samuel "On the speed of gravity and the Jupiter/Quasar measurement"}

\author{Sergei M. Kopeikin}

\address{Department of Physics and Astronomy, University of Missouri-Columbia, \\Columbia, Missouri 65211, USA\\
kopeikins@missouri.edu}

\maketitle


\begin{abstract}
Recent review article by S. Samuel "On the speed of gravity and the Jupiter/Quasar measurement" published in the {\it International Journal of Modern Physics D}, {\bf 13} (2004) 1753,  provides the reader with a misleading "theory" of the relativistic time delay in general theory of relativity. Furthermore, it misquotes original publications by Kopeikin and Fomalont \& Kopeikin related to the measurement of the speed of gravity by VLBI. We summarize the general relativistic principles of the Lorentz-invariant theory of propagation of light in time-dependent gravitational field, derive Lorentz-invariant expression for the relativistic time delay, and finally explain why Samuel's "theory" is conceptually incorrect and confuses the speed of gravity with the speed of light.   
\end{abstract}

\keywords{general relativity; speed of gravity; speed of light; relativistic time delay}

\section{Introduction}

Exact mathematical solution of the light geodesic equation in time-dependent gravitational field of arbitrary moving bodies has been constructed in a series of our publications \cite{ksge,ks,km}. This solution predicts that a light particle (radio wave) is deflected by the gravitational field of the moving body from its retarded, with respect to observer, position. The retarded position of the light-ray deflecting body originates from the Lienard-Wiechert solution of the linearized Einstein equations and can be found by solving the retarded-time equation which is a null characteristic of the gravitational field. We proposed relativistic VLBI experiment to measure this effect of retardation of gravity by the field of moving Jupiter via observation of light bending from a quasar \cite{apjl} and successfully completed this experiment in September 2002. The experiment testing observational phase-referencing technique with several reference calibrators is described in \cite{pr1} and the results of the main experiment of September 2002 are published in Astrophysical Journal \cite{apj}.

Samuel's paper \cite{s2} is an attempt to review the results of the experiment by making use of a linear Lorentz transformation of the static Shapiro time delay. This approach is physically insufficient for conceptual analysis of the experiment which requires matching of the first and second order effects in the relativistic theory of the time delay based on the Lienard-Wiechert solution of the Einstein equations. This is the main reason for the erroneous conslusions about the nature of the experiment mispresented in \cite{s2}. In the present paper we outline the basic equations of the complete Lorentz-invariant theory of the time delay and show mistakes in Samuel's linearized "theory" \cite{s2} originating from his first publication \cite{s1}.

\section{Retardation of Gravity and the Lienard-Wiechert Potentials}

We denote the barycentric coordinates of the solar system as $x^\alpha=(x^0,x^i)$, where $x^0=ct$, $x^i={\bm x}$, and $c$ is the fundamental speed limit. The metric tensor $g_{\mu\nu}=\eta_{\mu\nu}+h_{\mu\nu}$, where $\eta_{\mu\nu}={\rm diag(-1,1,1,1)}$ is the Minkowski metric, and $h_{\alpha\beta}$ describes gravitational field of the solar system in the linearized post-Minkowskian \cite{dam} approximation\footnote{Greek indices run from 0 to 3. Roman indices run from 1 to 3. The Greek indices are rised and lowered with the Minkowski metric. Bold letters denote spatial vectors. Repeated indices mean 
the Einstein summation rule. Euclidean dot and cross products of two vectors are 
denoted as ${\bm a}\cdot{\bm b}$ and ${\bm a}\times{\bm b}$ respectively.}. We impose the harmonic gauge conditions $\partial_\nu h^{\mu\nu}-1/2\partial^\mu h=0$, where $\partial_\nu\equiv\partial/\partial x^\nu$ denotes a partial derivative with respect to coordinate $x^\nu$, and $h\equiv\eta^{\mu\nu}h_{\mu\nu}$.

Linearized Einstein equations in harmonic coordinates assume the following form
\begin{equation}
\label{ee}
\left(-\frac{1}{c^2}\frac{\partial^2}{\partial t^2}+\nabla^2\right)h^{\mu\nu}=-\frac{8\pi G}{c^4}\left(T^{\mu\nu}-\frac{1}{2}\eta^{\mu\nu}T\right)\;,
\end{equation}
where $\nabla\equiv\partial/\partial x^i$, $T^{\mu\nu}$ is the stress-energy tensor of matter generating the gravitational field, and $T\equiv\eta^{\mu\nu}T_{\mu\nu}$. In what follows, we take the stress-energy tensor in the standard form of moving point-like masses with Dirac's delta-functions as amtter's support \cite{ll}. In linear approximation the gravitational field of the solar system bodies moving along their orbits with acceleration can be found from Eq. (\ref{ee}) separately for each body. For Jupiter, the perturbation of the metric tensor, $h^{\mu\nu}$, is obtained as the Lienard-Wiechert solution of the Einstein equations (\ref{ee}) which yields 
\begin{equation}\label{1} 
h^{\mu\nu}=-\frac{2GM_J}{c^4}\frac{2u^\mu u^\nu+\eta^{\mu\nu}}{r_\alpha 
u^\alpha}\;.
\end{equation}
Here $M_J$ is the mass of
Jupiter, $u^\alpha=\gamma(s)(c, {\bm v}_J(s))$ is its four-velocity
with ${\bm v}_J(s)=d{\bm x}(s)/ds$,
$\gamma(s)=(1-v_J^2(s)/c^2)^{-1/2}$ and the null-cone distance,
$r^\alpha=x^\alpha-x^\alpha_J(s)$, and Jupiter's worldline
$x^\alpha_J(s)=(cs, {\bm x}_J(s))$ are functions of the retarded time $s$,
determined as a solution of null cone equation
$\eta_{\mu\nu}r^\mu r^\nu=0$, that is
\begin{equation}
\label{2}
s=t-\frac{1}{c}|{\bm x}-{\bm x}_J(s)|\;.
\end{equation}
Eq. (\ref{2}) describes the propagation of Jupiter's gravity field from Jupiter to the point ${\bm x}$ with the fundamental speed $c$ \footnote{Because Eq. (\ref{2}) is the null cone of the gravitational field of moving Jupiter we used \cite{apj,cqg} a symbol $c_g$ in there instead of $c$ to mark the parameter which we fit to observations in the data processing procedure.}. The letter $c$ used here is the speed of gravity and should not be confused with the speed of light (the speed of a radio wave from a quasar measured by VLBI) as it has been mistakenly done by Asada \cite{ass}.  This fundamental conclusion of general relativity can be tested experimentally by observing gravitational interaction of light from a background source (quasar, star) with the time-dependent field of moving Jupiter and/or other solar system bodies \cite{apjl}.

\section{Electromagnetic Phase}

Radio interferometry (VLBI) measures the phase $\varphi$ of the wave front coming from a quasar.  The phase is determined from the eikonal
equation for electromagentic field (the quasar radio wave) in the geometric optics approximation of Maxwell's equations \cite{ll}  
\begin{equation}\label{eik}
g^{\mu\nu}\partial_\mu\varphi\partial_\nu\varphi=0\;,
\end{equation}
where in the linear approximation
$g^{\mu\nu}=\eta^{\mu\nu}-h^{\mu\nu}$.  Solution of the eikonal equation
is obtained by iterations after substitution of the Lienard-Wiechert gravitational potentials, Eq. (\ref{1}), into Eq. (\ref{eik}) \cite{abgr} 
\begin{equation}
\label{3}
\varphi=\varphi_0+\frac{\nu}{c}\left[k_\alpha x^\alpha+\frac{2GM_J}{c^2}\left(k_\alpha 
u^\alpha\right)\ln\Phi\right]\;,
\end{equation}
where $\Phi\equiv -k_{\alpha} r^\alpha$, $\varphi_0$ is a constant, $\nu$ is the radio frequency, $k^\alpha=(1,{\bm k})$ is a
null vector associated with the propagating radio wave, and the unit vector ${\bm k}$ is directed from the quasar to VLBI station \footnote{Notice that vector $k^\alpha$ does not coincide with the null characteristic of the gravitational field given by the interval $x^\alpha-x^\alpha_J(s)$ (see Fig. 1).}. 
 
Solution for eikonal given by Eq. (\ref{3}) describes propagation of a plane electromagentic wave scattered on the time-dependent gravitational potential of a point-like mass (Jupiter) moving with time-dependent velocity. We abandoned the acceleration-dependent terms in Eq. (\ref{3}) because of their smallness. Influence of the acceleration-dependent terms has been analysed in \cite{ks}. We notice that Eq. (\ref{3}) is Lorentz-invariant because it consists of the products of four-dimensional vectors remaining invariant with respect to the Lorentz transformations of the both Maxwell and Einstein equations. In particular, the argument of the logarithmic fucntion is $\Phi\equiv -k_{\alpha} r^\alpha=r-{\bm k}\cdot{\bm r}$, where $r=|{\bm r}|=\sqrt{{\bm r}\cdot{\bm r}}$, ${\bm r}={\bm x}-{\bm x}_J(s)$, and depend on the retarded position of Jupiter, ${\bm x}_J(s)$, calculated at the retarded time $s$ defined by Eq. (\ref{2}) of the null characteristic of the gravitational field of Jupiter. Thus, precise interferometric measurement of phase $\varphi$ of the electromagnetic wave scattered by the gravitational field of a moving gravitating body (Jupiter) allows to determine the null characteristic of the gravitational field and measure the ultimate speed of propagation of gravity which is expected to be equal to the speed of light in vacuum. This is because the measurable phase $\varphi\sim\ln\Phi(s)$ is a logarithmic function of the retarded time $s$ as evident from Eq. (\ref{3}). 

It is important to notice that the null characteristic of the gravitational field connecting the point of observation, ${\bm x}$, and the retarded position of Jupiter, ${\bm x}_J(s)$, can not be confused with the null characteristic of the electromagnetic field propagating along the space-time direction defined by the null vector $k^\alpha$ because this vector is not parallel to the null vector $r^\alpha = x^\alpha-x^\alpha(s)$ (see Fig. 1). Unfortunately, the four-dimensional Minkowskian diagram of the experiment shown in Fig. 1, was not understood by Asada \cite{ass} and Samuel \cite{s1,s2} who confused the null characteristic of the quasar radio wave and that of the gravitational field of Jupiter.         

\section{The Lorentz-invariant Theory of Time Delay}

The Lorentz-invariant relativistic time delay equation, generalizing the
static Shapiro delay, can be obtained outright from
equation (\ref{3}). We note that the phase $\varphi$ of electromagnetic wave,
emitted at the point $x^\alpha_0=(ct_0,{\bm x}_0)$ and received at the
point $x^\alpha=(ct,{\bm x})$, remains constant along the wave's path
\cite{ll}. Indeed, if $\lambda$ is an affine parameter along
the path, one has for the phase's derivative
\begin{equation}
\label{phase}
\frac{d\varphi}{d\lambda}=\frac{\partial\varphi}{\partial x^\alpha}\frac{
dx^\alpha}{d\lambda}=0\;,
\end{equation}
due to the orthogonality of the light rays and their wave fronts.
Eq. (\ref{phase}) means that $\varphi\left(x^\alpha(\lambda)\right)={\rm const.}$ in accordance with our assertion.  Equating two values of the phase at the event of emission,
$x^\alpha_0$, and that of observation, $x^\alpha$, and separating the time and space coordinates
one obtains from (\ref{3})
\begin{equation}
\label{tde}
t-t_0=\frac{1}{c}{\bm k}\cdot\left({\bm x}-{\bm
x}_0\right)-\frac{2GM_J}{c^3}\frac{1-c^{-1}{\bm k}\cdot{\bm
v}_J}{\sqrt{1-v^2_J/c^2}}\ln\left(r-{\bm k}\cdot{\bm
r}\right)\;,
\end{equation}
where both the distance ${\bm
r}={\bm x}-{\bm x}_J(s)$, $r=|{\bm x}-{\bm x}_J(s)|$, and the retarded
time $s$ are defined by the gravity null-cone equation (\ref{2}). The time
delay (\ref{tde}) of light propagating through the gravitational field of an arbitrary moving body was derived first by Kopeikin and Sch\"afer
\cite{ks} who solved equations for light geodesics in the retarded
Lienard-Wiechert gravitational field of the body. Klioner \cite{kl} also obtained
this expression for the case of a uniformly moving body by making use of the Lorentz transformation of the 
Shapiro delay (that is, both the Maxwell and Einstein equations) from static to
moving frame. Samuel's paper \cite{s2} represents, in fact, rather convoluted linearized analogue of Klioner's painstaking calculation \cite{kl}.  

\section{Relationship Between Our Theory and Samuel's Calculations}

Let us approximate 
Eq.~(\ref{3}) by introducing two angles $\Theta$
and $\theta$ between vector ${\bm k}$ and unit
vectors ${\bm p}={\bm R}/R$ and ${\bm l}={\bm r}/r$ correspondingly, where vectors ${\bm
R}={\bm x}-{\bm x}_J(t)$ and ${\bm r}={\bm x}-{\bm
x}_J(s)$ connect 
the point of observation ${\bm x}\equiv{\bm x}(t)$ with the present, ${\bm x}_J(t)$, and retarded, ${\bm x}_J(s)$, positions of Jupiter, respectively. By definition $\cos\Theta={\bm k}\cdot{\bm p}$ and
$\cos\theta={\bm k}\cdot{\bm l}$. The product $\Phi\equiv -k_\alpha r^\alpha=r(1-\cos\theta)\simeq r\theta^2/2$
for small angles. Hence, the phase variation caused by space-time
difference $\delta x^\alpha=(c\delta t, \delta{\bm x})$ between two VLBI antennas is
\begin{equation}
\label{5}
\delta\varphi=\frac{\nu}{c}\left(k_\alpha\delta x^\alpha+\frac{4GM_J}{c^2}\frac{\delta\theta}{\theta}\right)\;,
\end{equation}
where $\delta\theta=-{\bm n}\cdot\delta{\bm x}/r$ with ${\bm n}={\bm
l}\times({\bm k}\times{\bm l})$ as the impact vector of the light
ray with respect to the retarded (due to the finite speed of gravity) position of Jupiter ${\bm x}_J(s)$.
Undetectable terms of order $v_J/c$ have been neglected. The quantity $\delta
t=t_2-t_1$ is the measurable VLBI time delay and $\delta{\bm x}={\bm x}_2-{\bm x}_1\equiv{\bm B}$
is a baseline between two VLBI stations. Since VLBI stations measure
the same wave front, $\delta\varphi=0$. Thus, for sufficiently small angle $\theta$ Eq. (\ref{5}) yileds
\begin{eqnarray}
\label{6a}\delta t&=&c^{-1}{\bm k}\cdot{\bm B}+\Delta\;,\\\nonumber\\\label{6}
\Delta&=&-\frac{4GM_J}{c^3r}\frac{{\bm n}\cdot{\bm B}}{\theta}\;,
\end{eqnarray}
where we neglected small terms of the second order of magnitude.
 
Eq. (\ref{6}) can be also derived from Eq. (\ref{tde}) \cite{cqg}. It exactly coincides with the leading order term in Eq. (3.9) of Samuel's paper \cite{s2} and allows to establish the following relationship between Samuel's notations for the Earth-Jupiter distance $R_{EJ}$ and the "observable angle" $\theta_{obs}$ \footnote{These quantities appear in Eq. (1.3) in Samuel's paper \cite{s2} along with the "distance of the closest approach" $\xi$ spontaneously without rigorous mathematical description so that their meaning is fuzzy.}, and our notations for the null-cone distance $r$ and the angle $\theta$. Specifically, neglecting the Earth finite-size effects in our calculations, one obtains $R_{EJ}=r$ and $\theta_{obs}=\theta$. In what follows, we prefer to keep on using our notations. 

Eq. (\ref{6}) is the excess time delay caused by the scattering of the radio wave from the quasar on the gravitational Lienard-Wiechert potential of {\it moving} Jupiter. Though the terms depending on Jupiter's velocity do not show up in Eq. (\ref{6}) explicitly, they are surely incorporated in it {\it implicitly} through the retarded position of Jupiter, ${\bm x}_J(s)$, entering both the distance $r$, the angle $\theta$, and the impact parameter vector ${\bm n}$. Such implicit dependence of the relativistic expressions on velocity of gravitating body is typical for the, so-called, post-Minkowskian approximation scheme of general relativity operating with the retarded Lienard-Wiechert solutions of the Einstein equations \cite{dam,bel}. Thus, Eq. (\ref{6}) describes dynamical situation since (because of the orbital motion of Jupiter) the retarded quantities, $r$, $\theta$, and ${\bm n}$,  are not reduced to their static counterparts after their post-Newtonian expansion around the time of observation $t$. 
  
The post-Newtonian expansion of the retarded position of Jupiter, ${\bm x}_J(s)$, in Eq. (\ref{6}) around time of photon's arrival to observer, $t$, yields 
\begin{equation}
\label{yu}
{\bm x}_J(s)={\bm x}_J(t)+{\bm v}_J(t)(s-t)+O(s-t)^2\;.
\end{equation}
The difference $s-t$ is calculated by solving the gravity null-cone Eq. (\ref{2}). Substituting this solution to Eq. (\ref{yu}) yields
\begin{equation}
\label{wa}
{\bm x}_J(s)={\bm x}_J(t)-\frac{1}{c}{\bm v}_J(t)R+O(s-t)^2\;,
\end{equation}
where $R={\bm R}|$, and ${\bm R}={\bm x}-{\bm x}_J(t)$ is the vector lying on the hypersurface of constant time $t$ and connecting the point of observation, ${\bm x}$, and the present position of Jupiter, ${\bm x}_J(t)$. 

The post-Newtonian expansion given in Eq. (\ref{wa}) originates from the retarded nature of the Lienard-Wiechert gravitational potentials and, thus, describes the effect of the retardation of gravity \cite{apjl}. The presence of the retardation of gravity effect through the retarded position of Jupiter in the Lorentz-invariant Eq. (\ref{3}) and its small-angle approximation Eq. (\ref{6}) reflects the causal property of gravitational field which is a consequence of its finite speed. Causality and Lorentz-invariance of the gravitational field are tightly connected fundamental concepts and the measurement of the causal nature (retardation) of gravity in the Fomalont-Kopeikin experiment is equivalent to the proof that the gravitational field is Lorentz-invariant and vice versa. 

Samuel also uses the post-Newtonian expansion of the retarded coordinate of Jupiter in Eq. (5.3) of his paper \cite{s2}. He believes that the post-Newtonian expansion "arises because the position of Jupiter changes as the quasar signals travel from the Jupiter region to Earth". Jupiter does move as the quasar signal travels towards observer but the distance traveled by Jupiter is proportional not to the difference between the time $t^*$ of the closest approach of the quasar signal to Jupiter and the time of observation, that is $t^*-t$, but to the difference between the retarded time $s$ and the time of observation $t$, that is $s-t$, as clearly follows from Eq. (\ref{yu}). Therefore, the fuzzy concept of "the Jupiter region", which is repeatedly used by Samuel without rigorous mathematical definition, is, in effect, the retarded position of Jupiter defined by the Lienard-Wiechert solution of the gravity field equations. This consideration makes it evident that the origin of the post-Newtonian expansion in Eqs. (5.3)--(5.6) of Samuel's paper \cite{s2} is caused by the speed of gravity which propagates from moving Jupiter towards observer as well as the quasar signal does. This point was emphasized in our papers \cite{apjl,abgr}. Samuel overlooked the gravitational physics of the Jupiter-quasar experiment because of his approximate and, hence, insufficient solution of the problem of propagation of light rays in time-dependent gravitational fields. He was able to integrate the light-ray geodesics only for the case of the small-angle approximation ($\theta\ll 1$) when the time of the closest approach, $t^*$, of the quasar signal to Jupiter is comparable with the retarded time $s$ along the null characteristic of Jupiter's non-stationary gravitational field. It is for this reason that Samuel confused the time $t^*$ and the retarded time $s$ and replaced the concept of the propagation of gravity from the retarded position of Jupiter by the concept of the propagation of light from "the Jupiter region" \footnote{Similar mistake has been done also by Will in \cite{will} who used insufficiently elaborated $c_g$-parametrization of the Einstein gravity field equations.}.    

Substitution of the post-Newtonian expansion (\ref{wa}) to Eq. (\ref{6}) yields
\begin{equation}
\label{p1}
\Delta =\Delta_S+\Delta_R\;.
\end{equation} 
Here 
\begin{equation}
\label{p2}
\Delta_S=-(4GM_J/c^3\Theta)({\bm N}\cdot{\bm B})\;
\end{equation}
 is the static Shapiro time delay caused by Jupiter's gravitational field at the time of observation, where ${\bm N}={\bm p}\times({\bm k}\times{\bm p})$ is the impact vector of the light ray with respect to the present position of Jupiter ${\bm x}_J(t)$, and  
\begin{equation}
\label{7}
\Delta_R =\frac{4GM_J}{c^4R\Theta^2}\biggl[2
({\bm N}\cdot{\bm v}_J){\bm N}-({\bm K}\times({\bm v}_J\times{\bm K})\biggr]\;,
\end{equation}
is the post-Newtonian correction to the Shapiro time delay due to the the finite speed of gravity in the gravity null-cone equation (\ref{2}).
Eq. (\ref{7}) is the same as Eq. (4) from \cite{apj}. It describes the first post-Newtonian $v_J/c$ correction to $\Delta_S$ and can be detected because of the amplifying factor $\sim 1/\Theta^2$.
Samuel's Eq. (5.6) is an approximate form of our Eq. (\ref{7}). We notice that Samuel's notation for the angle $\theta_1\equiv\Theta$ in our notations. Physical origin of the post-Newtonian correction $\Delta_R$ to the static Shapiro time delay $\Delta_S$ is due to the Lorentz-invariant nature of the gravitational field caused by its finite speed of gravity as follows from Eq. (\ref{yu}).     

In his first paper \cite{s1} Samuel incorrectly assumed that the experiment directly compared the
radio position of the quasar with the optical position of Jupiter, and that the
direction of Jupiter was determined by "sunlight that has been reflected off of Jupiter" (see the second paragraph in section III of Samuel's paper \cite{s1} describing figure 1 which is similar with figure 2 of Samuel's paper \cite{s2}). This assumption would correspond to direct measurement of the
angle $\theta$ and hence no $v_J/c$ terms would be observed since they are not evident in Eq. (\ref{6}). This explains why Samuel has erroneously decided "that the $v/c$ effects are too small to have been measured in the recent experiment involving Jupiter and quasar J0842+1845" \cite{s1}. 
The
experiment, however, monitored the position of the quasar as a
function of the atomic time by the arrival of the quasar's photons at the
telescope, while the Jupiter's position entering the time delay Eq. (\ref{tde}) was determined separately by fitting VLBI data for the quasar to a
precise JPL ephemeris, evaluated at the same atomic time as the arrival of
a photon via standard transformations from ephemeris time to atomic
time \cite{apj}. The result of our fitting procedure was that Jupiter deflects light from its retarded position ${\bm x}_J(s)$ but not from its present position ${\bm x}_J(t)$. Thus, the difference $\Delta-\Delta_S$ was measured and the $v_J/c$ correction $\Delta_R$ was determined within precision of 20\% \cite{apj}. The measurement of the post-Newtonian correction (\ref{7}) to the Shapiro time delay (\ref{p2}) is direct demonstration that gravity does propagate with the same speed as the speed of light. Samuel's claim that "Fomalont and Kopeikin's announcement that the speed of gravity is the speed of light to within 20\% has no content" is based on his inability to distinguish between gravitational and electromagnetic effects in the post-Newtonian expansion of the Lorentz-invariant time delay Eq. (\ref{tde}) predicting that any moving body deflects light by its gravitational field from the retarded position in accordance with the causal (Lorentz-invariant) nature of gravity.

\section{Discussion and Further Particular Comments}

\subsection{On the Ideology of the Experiment}

The paper by Samuel \cite{s2} is an attempt to protect his misleading calculations \cite{s1} published in \cite{s1} which conceptual inconsistency was revealed in \cite{apj,cqg,abgr,vc}. Unfortunately, it does not provide either new mathematical details or more deep insight to the problem. By making use of a linearized Lorentz transform of a spherically-symmetric gravitational field Samuel succeeds in calculation of the first few terms of the differential Shapiro time delay caused by moving Jupiter that approximates the original result by Kopeikin \cite{apjl} (see Eq. (\ref{tde})) given in terms of the retarded time, $s$, connecting the point of observation, ${\bm x}\equiv{\bm x}(t)$, and the retarded position of a light-ray deflecting body (Jupiter), ${\bm x}_J(s)$. Our derivation of Eq. (\ref{tde}) makes it clear that the electromagnetic signal from a quasar is observed at the time, $t$, and Jupiter deflects it at the retarded time, $s=t-r/c$, where $r=|x(t)-x_J(s)|$ is the radial coordinate of Jupiter with respect to observer directed along the null characteristic of the gravitational field defined by the null cone equation $\eta_{\mu\nu}r^\mu r^\nu=0$. Both the retarded coordinate of Jupiter, ${\bm x}_J(s)$, and the retarded time, $s$, originate from the Lienard-Wiechert solution of the linearized Einstein equations which is a hyperbolic (wave-type) D'Alembert equation. Lorentz-invariant theory of the propagation of light rays through the time-dependent field of the gravitational retarded potentials reveals that the light particle (photon) is deflected by moving Jupiter when it is located at the retarded position ${\bm x}_J(s)$ due to the finite speed of gravity. Experimental testing whether Jupiter deflects light from its orbital position taken at the retarded time $s$ due to the finite speed of gravity or at the time of observation $t$, is a direct probe of the numerical value of the speed of gravity which must be equal to the speed of light according to Einstein. This is the key idea of the experiment which has been put forward in our publication \cite{apjl} and practically tested in September 2002 \cite{apj}. Unfortunately, the einsteinian gravitational physics of the experiment is greatly misunderstood and conceptually misrepresented both in Samuel's papers \cite{s1,s2} and in \cite{ass,will,car} \footnote{Critical discussion of the formally correct, but conceptually misleading point of view on the gravitational physics of the process of light scattering by the gravitational field of a moving body presented in \cite{car}, requires more elaborated mathematical technique than that presented in the present paper, and will be given somewhere else. Some details are available in \cite{mw}.}. 

\subsection{What Was Observed and Tested in the Experiment}

Samuel prefers to re-express our original result for time delay (\ref{tde}) in terms of the "observable angle" $\theta_{obs}\equiv\theta$ in terms of "the distance of the closest approach $\xi$".  The vertex of this angle is at the point of observation, ${\bm x}(t)$, and it has two legs -- one leg is directed in the sky towards the quasar and another one is directed towards retarded position of Jupiter, ${\bm x}_J(s)$. If both Jupiter and the quasar were observed simultaneously in radio or in optics, then, the legs composing the "observable angle" $\theta$ would be formed by the null characteristics of electromagnetic signals emanating correspondingly from the quasar and from Jupiter to the observer. However, Jupiter was not observed by VLBI in the Fomalont-Kopeikin experiment at all, only the quasar was - so that only one leg of "the observable angle" is the null characteristic of electromagentic field. Jupiter affected the propagation of the quasar radio wave by its gravitational field and the effective position of Jupiter in the sky (the second leg of "the observable angle") was determined in the process of the data analysis of the residual phase of the quasar radio wave.  The time delay Eq. (\ref{tde}) tells us that the direction in the sky connecting the observer and Jupiter is the null characteristic of the gravitational field of moving Jupiter. Thus, Samuel's "observable angle" $\theta_{obs}\equiv\theta$ is made of the null lines one of which belongs to the electromagnetic field (radio wave from the quasar) but another one belongs to the gravitational field of moving Jupiter, which deflects the radio signal of the quasar not instantaneously but with the retardation to comply with the Lorentz-invariant symmetry of the Einstein equations \footnote{See Fig. 1 for graphical illustration of this point.}.  

The goal of our measurement was to distinguish between the two hypothesis: 
\begin{itemize}
\item[(1)] Einsteinian gravity. Jupiter deflects light by its gravitational field from its retarded position ${\bm x}_J(s)$ as shown in our time-delay Eq. (\ref{tde});
\item[(2)] Newtonian gravity. Jupiter deflects light by its gravitational field instantaneously and its position in the time delay Eq. (\ref{tde}) must be taken at the time of observation, $t$. 
\end{itemize}
In effect, we did not measure any angles in the sky directly during the time of observation as Samuel seems to believe. What was measured is the residual phase $\delta\varphi$ of the quasar radio wave, allowing us to determine the difference between times of arrival of the same front of the quasar radio wave to two VLBI stations \cite{apj}. This difference depends on the position of Jupiter on its orbit and our goal was to prove that the observed differential time delay is affected by the gravitational field of Jupiter acting from its retarded position, ${\bm x}_J(s)$, but not from its position, ${\bm x}_J(t)$, taken at the time of observation. The difference between the two hypothesis gives rise to the post-Newtonian correction Eq. (\ref{7}) to the Shapiro time delay (\ref{p2}) which had to be zero if the speed of gravity would be infinite. We proved experimentally \cite{apj} that the first hypothesis is correct within 20\% (that is the post-Newtonian correction $\Delta_R\not=0$) which means that the speed of gravity has the same value as the speed of light.

Samuel misinterprets the physical origin of the observed retardation-of-gravity effect in the time delay of light because of his insufficiently elaborated mathematical solution of the problem of light propagation in the time-dependent gravitational field of a moving body. While our approach starts from the basic principles and solution of the gravity field equations, Samuel's development deals only with the final product of the theory -- the time delay Eq. (\ref{6a}), which incorporates effects of the propagation of both gravitational and electromagentic fields. When making Lorentz transformation of the time delay Eq. (\ref{6a}) Samuel has missed the point that this transformation transforms not only the electromagnetic but the gravitational field as well. Thus, the experiment effectively measures the discrepancy between the Lorentz-invariant symmetry of the gravitational field with respect to the Lorentz transformation symmetry of the electromagnetic field which could arise due to the possible difference between the speed of gravity and light. The experiment has confirmed the Lorentz invariance of the gravitational field because no difference between the speeds of gravity and light was found within 20\% \cite{apj}. 

\subsection{On the Relativistic Jargon and the Symbol $c$}
    
Einstein's theory of general relativity predicts that an electromagnetic wave is deflected by the gravitational field of moving body (Jupiter) from its retarded position, ${\bm x}_J(s)$, taken at the retarded instant of time $s=t-r/c$, where $c$ is conceptually the fundamental speed of propagation of gravitational field because it enters the Lienard-Wiechert gravitational potentials. For historical reasons, it is customary to call the symbol $c$ as "the speed of light". We emphasize however that this term is misleading when one considers the Lienard-Wiechert solution of the Einstein gravity field equations since in this case $c$ is the speed of gravity characterizing the degree of opening of the null cone made up of the characteristics of the gravitational field. This interpetation of the symbol $c$ entering the retarded coordinate of a moving gravitating body remains true in the case of the physical process of scattering of electromagnetic wave on the Lienard-Wiechert gravitational potential of this body. This point was misinterpeted by Asada \cite{ass}, Will \cite{will}, and Carlip \cite{car}. Samuel \cite{s2,s1} seems to be also confused with the existing conventional terminology and is about to interpret any physical effect depending on $c$ as associated with propagation of light because $c$ is always "the speed of light" by his definition 
\footnote{The same jargon is used by Asada \cite{ass} and Will \cite{will} 
that prevents them to see the true nature of the retardation of gravity effect measured in the Fomalont-Kopeikin experiment \cite{apj}.}. 
Einstein's gravity field equations contain "the speed of light" constant $c$ but it does not mean that the measurement of this fundamental constant for gravitational field (as it is done in \cite{apj}) is reduced (or equivalent) to the measurement of the physical speed of light even if the light is used for such measurement. Various fasets of the symbol $c$ and the pitfalls related to the physical interpetations of various experiments have been discussed recently in \cite{cqg,uzan}.

\subsection{On the Point of the Closest Approach}

Samuel's calculations are also based on the assumption that an electromagnetic signal is effectively deflected at the point of its closest approach to the moving Jupiter \footnote{The same assumption was also used by Will \cite{will}}. From the first glance this assumption looks plausible but turns out to be generally incorrect in exact mathematical solution of the problem shown in the present paper as well as in our previous works \cite{ksge,ks,km,apjl,abgr,pla} and the work of Klioner \cite{kl}. First of all, gravitational field is long-ranged and the process of gravitational light deflection (electromagentic wave scattering) does not take place in one event. Relativistic time delay of light is the integral effect and, strictly speaking, it can be expressed in terms of the impact parameter of the light ray, taken at the time of the closest approach $t^*$, only approximately (see \cite{ksge,ks,km} for more detail). But even in the case of the approximation the point of the closest approach of light to the moving body loses its physical content when the Lorentz transformation is applied to transform the time delay from the static to a moving frame. Advanced calculation (missed in \cite{s2,s1}) makes it evident that the integral effect of the light-ray time delay is coming out from the retarded position of the light-ray deflecting body \cite{ks,km,apjl} which is in accordance with the Lienard-Wiechert solution of the Einstein gravity field equations describing propagation of the gravitational field \footnote{We emphasize once again that there is no light propagating from the retarded position of the light-ray deflecting body in the time delay Eq. (\ref{tde}) to observer as  it was erroneously deduced by Asada \cite{ass}, it is gravity which propagates.}. 

Recently Klioner \cite{kl} has calculated the deflection and the time delay of light in the gravitational field of a uniformly moving body by making use of the Lorentz transformation technique. In contrast to the linearized Samuel's "theory" \cite{s2,s1}, Klioner's calculation takes into account all velocity-dependent terms in the Lorentz transformation which significantly supersedes Samuel's consideration \cite{s2,s1}. Klioner's paper \cite{kl} delivers an independent proof that Jupiter must deflect light from its retarded position ${\bm x}_J(s)$ in accordance with the Lienard-Wiechert solution of the gravity field equations \cite{ks,km,apjl,cqg}. The effect of the retardation of gravity in the case of a uniformly moving body is equivalent to the property of the gravitational field to be Lorentz-invariant and to propagate with the fundamental speed $c$ which, at the same time, is the parameter of the Lorentz transformation of the Einstein gravity field equations. Our experiment has tested this property of gravity by observing that Jupiter's gravitational field deflects the quasar radio wave out of Jupiter's retarded position lying on the null characteristic of the gravitational field which connects the VLBI station and Jupiter and is given by the retarded time Eq. (\ref{2}).

Different mathematical technique \cite{ksge,ks,km,pla} also demonstrates that the point of the closest approach of photon to the moving body plays no physical role because the time of the closest approach, $t*$, drops out of the final equation for the light bending and/or relativistic time delay. What matters is the retarded time $s=t-r/c$ of the Lienard-Wiechert solution of Einstein's equations, where $c$ is the speed of propagation of gravity from the moving body to the point of observation. The time of the closest approach, $t*$, approximates the retarded time $s=t-r/c$, if the impact parameter of the light ray to the moving body is small enough (see \cite{ksge,apjl} for more detail). This mathematical approximation is physically possible because the gravitational field propagates with the same speed as light in general theory of relativity. In fact, Samuel \cite{s1,s2} unknowingly  used this approximation to replace the retarded time $s$ with the time of the closest approach $t^*$ which has led him to the confusion of the propagation of gravity effect with that of the propagation of the quasar signal. His statement that the "Shapiro time delay difference is due to relatively {\it short-distance effects} is correct but these {\it short-distance effects} occur near the retarded position of Jupiter ${\bm x}_J(s)$ taken at the retarded time $s$ while the time of the closest approach $t^*$ is an approximation which, if it is used improperly, simply misinterpets the gravitational physics of the Jupiter-quasar experiment.

\subsection{On the Minkowski Diagram of the Jupiter-Quasar Experiment}

Samuel's graphic analysis of the Fomalont-Kopeikin experiment is shown in figures 1 and 2 of the paper \cite{s2}. Unfortunately they do not grasp the relativistic spirit of the Jupiter-quasar experiment. The Minkowski diagram is much more adequate for picking up the idea of the measurement of the speed of gravity with VLBI. Analysis of the experiment in terms of the Minkowski diagrams is given by Kopeikin in the paper \cite{cqg} and in the proceedings of the 14th Midwest Relativity Meeting \cite{mw}. Minkowski diagram of the experiment shown in Fig. 1 of the present paper clearly demonstrates the origin of the observed retarded position of Jupiter as caused by the Lorentz-invariant nature of the gravitational field and its finite speed of propagation. A null direction connecting observer and Jupiter in the relativistic time delay Eq. (\ref{tde}) is that corresponding to the null characteristic of the gravitational field. This is because of two facts: (1) it originates from the retarded Lienard-Wiechert solution of the gravity field equations, and (2) there is no radio or light wave  emitted or reflected by and propagating from Jupiter towards observer during the time of the experiment. Thus, the measurement of the direction of the null characteristic of the gravitational field through the retarded coordinate of Jupiter, ${\bm x}_J(s)$, through the relativistic time delay is a direct confirmation of Einstein's prediction that gravitational field has the same speed of propagation as the speed of light. The reader should not interpret the retarded position of Jupiter shown in figure 2 of Samuel's paper \cite{s2} as caused by propagation of sunlight reflected from Jupiter as Samuel \cite{s1,s2} erroneously used to believe. Unfortunately, this was either overlooked or ignored by referees of papers \cite{s1,ass} despite of our persistent attempts to draw their attention to this, inconsistent with observations, fact. As explained in \cite{apj,cqg} the radio emission of Jupiter or the sunlight reflected by Jupiter can not be, and was not, observed at any VLBI station due to the specific technical limitations of VLBI. Anyone who discusses the nature of the Jupiter-quasar experiment must learn how VLBI operates and detect radio signals before making any statements about the nature of the measured effect of the retardation of gravity. The retarded position of Jupiter, measured in the experiment through the best fit of the observed VLBI time delay to its theoretical value, is due to the finite speed of gravity and reflects the fundamental fact that the gravitational field is Lorentz-invariant and its null characteristics coincide with the null characteristics of the electromagnetic field withing 20\% \cite{apj}. 

\subsection{On the speed of gravity parameter $c_g$}

Einstein's theory of general relativity does not require introduction of any parameter for the speed of gravity $c_g$ because it is Lorentz invariant and the speed of gravity is equal to the speed of light. All general-relativistic effects associated with the speed of gravity can be easily and unambiguously identified through the retarded positions of the gravitating bodies in the post-Minkowskian solutions of the gravity field equations and equations of motion. Will \cite{book} has decided to introduce the parameter $c_g$ for the speed of gravity to facilitate discussion of the relativistic effects associated with the speed of gravity from those caused by the "speed of light" $c$. Introduction of $c_g$ to the Einstein equations is not trivial and requires to retain all differential relationships of general relativity for any value of $c_g$. This task was not fulfilled in \cite{book}. 

We have found \cite{cqg} the $c_g$-parametrization of Einstein's equations which preserves all differential and algebraic properties of those equations for any value of the parameter $c_g$. This $c_g$-parametrization requires introduction of a global unit vector field, $V^\alpha$, which goal is to keep the Lorentz invariance of the gravity field equations. The difference between $c_g$ and $c$ leads to one kind of observable effects while existence of spatial components, ${\bm V}$, of the vector field $V^\alpha$ leads to apperance of the preferred frame effects (the PPN parameters $\alpha_1$, $\alpha_2$) which have been strongly limited by pulsar timing and lunar laser ranging observations \cite{book}. Hence, in the paper \cite{cqg} we did not analyzed the preferred frame effects caused by ${\bm V}$ and worked in the coordinate system, where $V^\alpha=(1,0,0,0)$, to concentrate on the discussion of the retardation of gravity effect \cite{apj}. Therefore, spatial vector field ${\bm V}$ does not appear explicitly in our derivation of the relativistic time delay equation. However, correct transformation of our time delay equation, derived in \cite{cqg} for the case of $c_g\not=c$, from one frame to another requires accounting for the preferred frame vector filed ${\bm V}$, which was not done by Samuel in deriving his equation (7.2) in \cite{s2}. Thus, Samuel's criticism of my paper \cite{cqg} is completely unfounded and is based on his misunderstanding of the mathemtical formalism which has been worked out in \cite{cqg}.

\begin{figure*}
\includegraphics*[height=16cm,width=14cm]{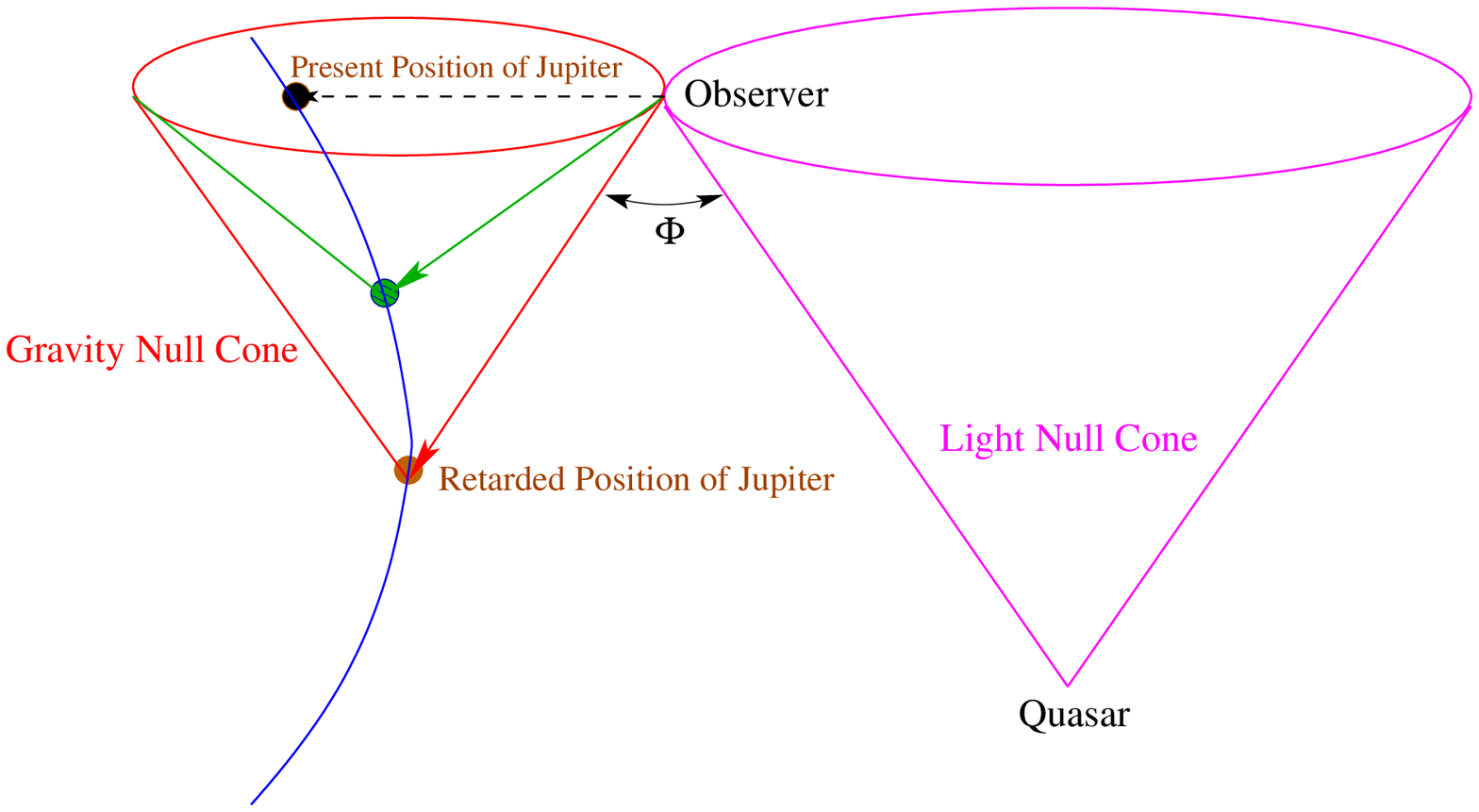}\noindent
\caption{\label{fig-1}Two null cones related to the experiment are shown. The gravity field cone describes the retarded Lineard-Wiechert solution of the Einstein equations and reflects the Lorentz invariant nature of the gravitational field. The light cone shows propagation of light from the quasar. The relativistic perturbation of a light ray measured by an observer takes place when the gravity cone of Jupiter passes through the observer. The VLBI experiment measures the Minkowski dot product $\Phi$ between two null vectors at the point of observation directed to the quasar and to Jupiter respectively. Had Jupiter not been detected at the retarded position on its world line the speed of gravity were not equal to the speed of light, and the Einstein principle of relativity for gravitational field would be violated. The experiment did not find any violation of the Lorentz-invariance of the gravitational field and confirmed that the speed of gravity equals to the speed of light within 20\%.}
\end{figure*}
\end{document}